\documentclass{article}

\usepackage{makeidx}         % allows index generation
\usepackage{graphicx}        % standard LaTeX graphics tool
\usepackage{epsfig}                             
        
\usepackage{subfig}        
        
\usepackage{cite}        
                             
\usepackage{multicol}        % used for the two-column index
\usepackage[bottom]{footmisc}% places footnotes at page bottom

\usepackage{newtxtext}       %
\usepackage{newtxmath}       % selects Times Roman as basic font

\usepackage{url} % Use the package "url.sty" to avoid problems with special characters used in your e-mail or web address

\begin{document}

\title{Topological Data Analysis of Spatial Systems}
\author{Michelle Feng \\ California Institute of Technology \\ \\ Abigail Hickok \\ University of California, Los Angeles \\ \\ Mason A. Porter \\ University of California, Los Angeles}

%\author{Michelle Feng, Abigail Hickok, and Mason A. Porter}

%\institute{Michelle Feng \at Department of Computing + Mathematical Sciences, California Institute of Technology, Pasadena, CA 91125, USA \email{mfeng@caltech.edu}
%\and Abigail Hickok \at Department of Mathematics, University of California, Los Angeles, CA 90095, USA \email{ahickok@math.ucla.edu}
%\and Mason A. Porter \at Department of Mathematics, University of California, Los Angeles, CA 90095, USA \email{mason@math.ucla.edu}}

\maketitle

%\abstract*{In this chapter, we discuss applications of topological data analysis (TDA) to spatial systems. We briefly review the recently proposed level-set construction of filtered simplicial complexes, and we then examine persistent homology in two cases studies: street networks in Shanghai and hotspots of COVID-19 infections. We then summarize our results and provide an outlook on TDA in spatial systems.}

\abstract{In this chapter, we discuss applications of topological data analysis (TDA) to spatial systems. We briefly review the recently proposed level-set construction of filtered simplicial complexes, and we then examine persistent homology in two cases studies: street networks in Shanghai and hotspots of COVID-19 infections. We then summarize our results and provide an outlook on TDA in spatial systems.}

%%%%%

%%%%%

\section{Introduction}
\label{sec11}

To improve our understanding of spatial systems, it is important to develop methods that directly probe the effects of space on their structure and dynamics. Many complex systems have a natural embedding in a low-dimensional space or are otherwise influenced by space \cite{barthelemy2018,newman2018}. Spatial effects significantly influence both their structure and their dynamics. One way to gain information about the global structure of spatial systems is by studying notions of `connectness', `holes', and `cavities'. Consequently, it is not surprising that many researchers have used topological data analysis (TDA), usually in the form of persistent homology (PH), to study a diverse variety of spatial systems. For example, TDA has been used to study granular and particulate systems \cite{lia2018,Buchet2018}, neuronal networks \cite{sizemore2019}, leaf-venation patterns \cite{leaf_optimization}, networks of blood vessels \cite{Byrne2019}, aggregation models \cite{topaz2015}, spatial percolation \cite{speidel2018}, human migration \cite{Ignacio2019}, voting patterns \cite{feng2021}.

Analyzing PH allows one to quantify holes in data in a meaningful way and has made it possible to apply homological ideas to a wide variety of empirical data sets \cite{otter2017}. To study PH, one needs to construct a filtered simplicial complex. (See, e.g., Chapter 3 and \cite{gunnar-natphys}.) In \cite{feng2021}, Feng and Porter developed new types of filtered simplicial complexes that incorporate spatial information. In \cite{feng2020}, they applied their new constructions to synthetic spatial networks, city street networks, spiderwebs, and snowflakes. Other studies have also incorporated spatial information into PH (see, e.g., \cite{ron2015,Kanari2016QuantifyingTI,Byrne2019}). Recently, researchers have also extended TDA methods other than PH --- ones that use persistent cohomology \cite{stolz2020} and the Euler characteristic \cite{smith2021} --- to incorporate spatial information. 

In the present chapter, we discuss two case studies of PH to spatial systems. We use a level-set construction of simplicial complexes, which were introduced recently in \cite{feng2021}, to (1) city street networks in Shanghai\footnote{This case study is related to an example in \cite{feng2020}.} and (2) hotspots of COVID-19 infections. Through these examples, we illustrate the importance of incorporating spatial information when doing TDA on spatial systems.

Our chapter proceeds as follows. In Section \ref{sec2}, we discuss the level-set construction of filtered simplicial complexes. We use these complexes to study PH for city street networks in Shanghai in Section \ref{sec3} and for hotspots of COVID-19 infections in Section \ref{sec4}. In Section \ref{sec5}, we conclude and give a brief outlook on TDA in spatial systems.

%%%

\section{Level-Set Complexes}
\label{sec2}

We now briefly review the level-set construction of filtered simplicial complexes that was introduced recently in \cite{feng2021}. For discussions of other types of filtered simplicial complexes (which are often called simply ``filtrations''), see Chapter 3 and \cite{otter2017}.

In a level-set filtration, one describes data as a manifold. Let $M$ denote a two-dimensional (2D) manifold, such as data in an image format. We construct a sequence
\begin{equation*}
	M_0 \subseteq M_1 \subseteq \cdots \subseteq M_n
\end{equation*}	
of manifolds (where $M_0$ is an approximation of $M$) as follows. At each time $t$, we evolve the boundary $\Gamma_t$ of $M_t$ outward according to the level-set equation of front propagation \cite{osher2003}. Specifically, for a manifold $M$ that is embedded in $\mathbb{R}^2$, we define a function $\phi(\vec{x},t)\!: \mathbb{R}^2 \times \mathbb{R} \to \mathbb{R}$, where $\phi(\vec{x},t)$ is the signed distance function from $\vec{x}$ to $\Gamma_t$ at time $t \geq 0$. We propagate $\Gamma_t$ outward at velocity $v$ using the partial differential equation
\begin{equation} \label{level}
	\frac{\partial \phi}{\partial t} = v |\nabla \phi |
\end{equation}
until all homological features die. The evolution \eqref{level} gives a signed distance function at each time $t$. We take $M_{t}$ to be the set of points $\vec{x}$ such that $\phi(\vec{x},t) > 0$. (This corresponds to points inside the boundary $\Gamma_t$.) In our examples in this chapter, we use $v=1$.

By imposing $\{M_i\}$ over a triangular grid of points (see \cite{feng2021}), we obtain a corresponding simplicial complex $X_i$ for each $M_i$. Because the level-set equation \eqref{level} evolves outward, we satisfy that condition that $X_i \subseteq X_{i+1}$ for all $i$, so $\{X_i\}$ is a filtered simplicial complex.

%%%%

\section{Case Study: Street Networks in Shanghai}
\label{sec3}

This case study is an extended discussion of one of the examples in \cite{feng2020}. In this case study, we use level-set complexes to examine patterns in city street networks. We focus on the city of Shanghai, which has a long history of urban development \cite{yeung_sung}. The discussion in \cite{feng2020} used PHs of level-set complexes to classify a variety of small street networks from different neighborhoods of Shanghai. In the present discussion, however, we closely examine the PHs of street networks in several different neighborhoods of Shanghai. Computing PH (and, more generally, using TDA) allows us to detect both topological and geometric properties of city blocks in these neighborhoods. These properties may reflect differences in the development of city streets across time and cultural influences.

The city of Shanghai was first inhabited about 6000 years ago during China's Warring States period. Over the course of several millennia, Shanghai has experienced urban growth, with a variety of developmental styles, over many distinct time periods \cite{yeung_sung}.
These different architectural and urban-planning styles reflect a diversity of different views by the various powers of Shanghai for how the city should be structured.
In the following paragraphs, we use PH to highlight street networks in several distinct neighborhoods of Shanghai. We draw connections between the history of these neighborhoods and the topological features that we observe in their PHs.

\begin{figure}[htbp]
\centering
    \subfloat[Laoximen]{\includegraphics[width = .25\textwidth]{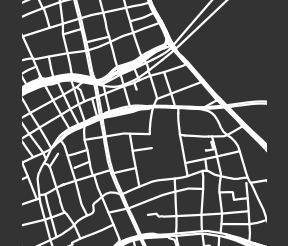}}
    \hspace{10mm}
    \subfloat[Former French concession]{\includegraphics[width = .25\textwidth]{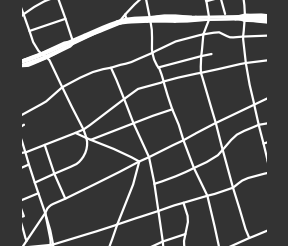}}
    \hspace{10mm}
    \subfloat[Pudong New Area]{\includegraphics[width = .25\textwidth]{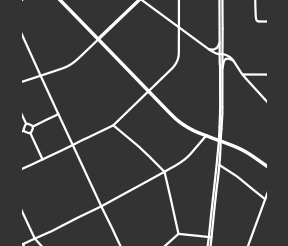}}
\caption{Street networks from three different neighborhoods of Shanghai. (We generated these maps using {\sc OSMnx} \cite{boeing_2017}.)
}
\label{fig:streetmaps}
\end{figure}

We use networks from {\sc OSMnx} \cite{boeing_2017} as input data. Our street networks are images of street maps; they consist of a 2 km by 2 km square that is centered at a given set of (latitude, longitude) coordinates. We show three such street maps in Figure~\ref{fig:streetmaps}. In Figure \ref{fig:streetmaps}(a), we show a street map from Laoximen (``Old West Gate''), a neighborhood that was built up around the western gate of Shanghai's original city walls. In Figure ~\ref{fig:streetmaps}(b), we show a street map from the former French concession, which was a French colonial territory from 1849 to 1943. In Figure \ref{fig:streetmaps}(c), we show a street map from Pudong New Area, which is a modern financial district that has developed mostly over the last few decades.

\begin{figure}[htbp]
\centering
    \subfloat[]{\includegraphics[width = .18\textwidth]{figures/shanghai/pudong-6.png}}
    \hspace{2mm}
    \subfloat[]{\includegraphics[width = .18\textwidth]{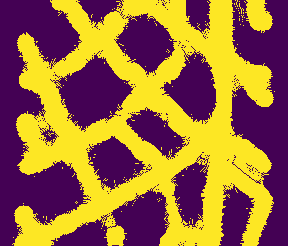}}
    \hspace{2mm}
    \subfloat[]{\includegraphics[width = .18\textwidth]{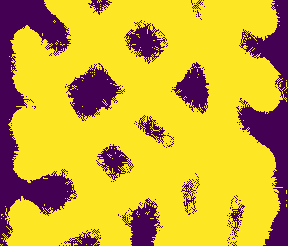}}
    \hspace{2mm}
    \subfloat[]{\includegraphics[width = .18\textwidth]{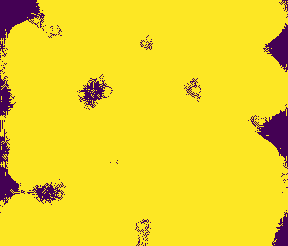}}
    \hspace{2mm}
    \subfloat[]{\includegraphics[width = .18\textwidth]{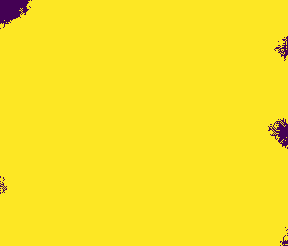}}
\caption{Selected steps of the level-set evolution of the map of Pudong New Area from Figure~\ref{fig:streetmaps}c. As the level-set complex evolves, the streets expand and fill in the blocks. Smaller blocks fill in faster.}
\label{fig:pudong_ls}
\end{figure}

From the street maps in Figure~\ref{fig:streetmaps}, we obtain level-set complexes following the approach in Section \ref{sec2}. In Figure~\ref{fig:pudong_ls}, we show the level-set complex that corresponds to the map in Figure~\ref{fig:streetmaps}(c). This level-set complex begins with line segments that represent the streets in the network. The streets expand outward as we add simplices to the complex. We thereby capture city blocks as homological features, whose death times increase as the sizes of the blocks increase. (Larger blocks take longer to be ``filled'' by the expanding streets in the simplicial complex.)

To visualize the features of the PHs of the street-map images, we use {\em persistence diagrams} (PDs). PDs represent homological features as points on a scatter plot. We plot each feature at a point $(b, d)$, where $b$ is the birth time of the feature and $d$ is its death time. We show zero-dimensional (0D) features in pink and one-dimensional (1D) features in blue. Because features cannot die before they are born, all points must lie on or above the identity line $g(x)=x$. More persistent features lie farther above this line. See Chapter 3 and \cite{otter2017} for more information about PDs.

\begin{figure}[htbp]
\centering
    \subfloat[Laoximen]{\includegraphics[width = .25\textwidth]{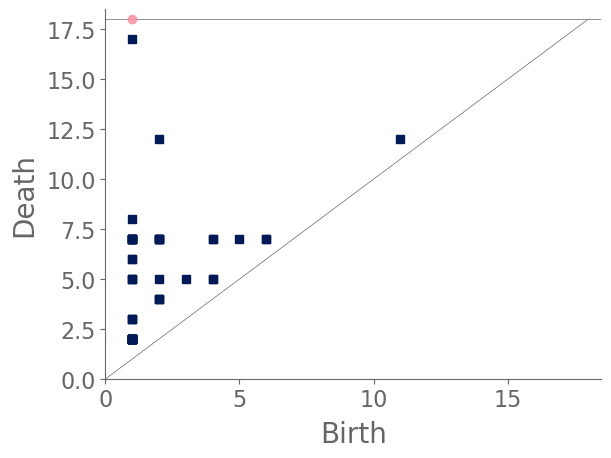}}
    \hspace{10mm}
    \subfloat[Former French concession]{\includegraphics[width = .25\textwidth]{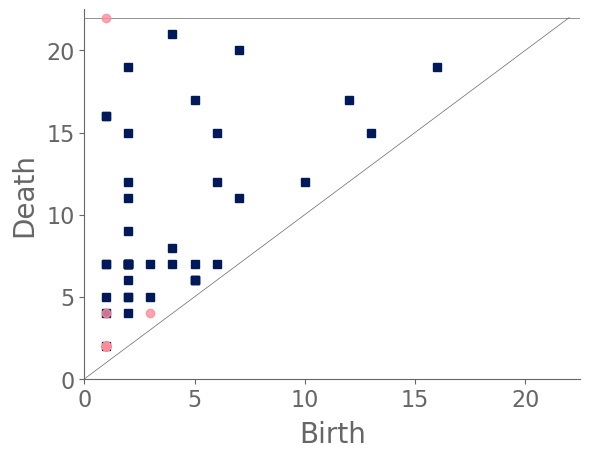}}
    \hspace{10mm}
    \subfloat[Pudong New Area]{\includegraphics[width = .25\textwidth]{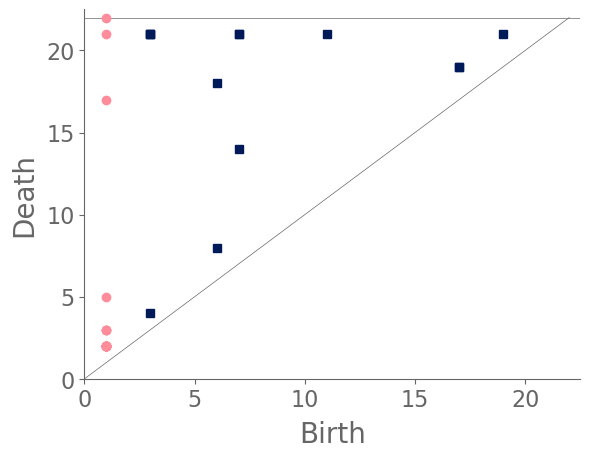}}
\caption{Persistence diagrams of street networks from three neighborhoods of Shanghai. By comparing the birth and death times of the features in the PDs, we observe differences in block-size distributions and block shapes in these neighborhoods. These differences reflect the different developmental histories of these three areas.
}
\label{fig:shanghai_pds}
\end{figure}

In Figure~\ref{fig:shanghai_pds}, we show the PDs that correspond to the maps in Figure~\ref{fig:streetmaps}. The PD of Laoximen [see Figure~\ref{fig:shanghai_pds}(a)] reveals that most of the 1D features have death times of less than 10. This indicates that the city blocks in this area are relatively small. Additionally, although many of the features of the map of Laoximen are born at early times (such features are close to the vertical axis of a PD), there are also several points close to the diagonal that have later birth times. These points correspond to features that tend to occur when a street map has dead ends. As the level-set complex evolves, dead ends expand. This can result in a single block being ``pinched'' into multiple smaller blocks when the dead end connects to the streets that border the block. Similarly, blocks that are not rectangular because of winding roads can be ``pinched'' into smaller blocks when narrower areas fill in faster than wider areas. In the street map of Laoximen, there are a large number of dead ends and winding streets. Street designs like these, which do not resemble rectangular grids, are less common in modern street layouts than in older ones \cite{Boeing2019}. We observe in Figure \ref{fig:streetmaps} that the southern part of our Laoximen map seems to contain more of these features than other parts of the map. This particular area of the map includes one of the oldest remaining neighborhoods of Laoximen.\footnote{This part of Laoximen has been slated for redevelopment since 2017 \cite{kanagaratnam2017}. When we obtained these street maps in 2019, residents were fighting redevelopment efforts and development had not yet begun \cite{walsh2019}. It remains to be seen how this part of our Laoximen street map will change as a result of redevelopment.}
Much of the area around it has been demolished and redeveloped.

The PD of the former French concession [see Figure~\ref{fig:shanghai_pds}(b)] has more features with death times between 10 and 20 than is the case for the PD of Laoximen. This indicates the existence of medium-size blocks, and we see in Figure \ref{fig:streetmaps} that the blocks in the former French concession are generally larger than those in Laoximen. We still observe many features with death times that are less than 10, so the street network of the former French concession does have a variety of block sizes. Although it has fewer dead ends than Laoximen, many of the blocks in the former French concession are not rectangular because of its curved roads. Like Laoximen, the former French concession has experienced much redevelopment in the last several decades \cite{guan1996}. However, many of the original buildings and streets remain, and the former French concession is a popular tourist destination because of its European-style buildings and streets. Extra-settlement roads that were built by the French colonial government, spacious residential lots, and its wide and tree-lined streets are reflected in its street map.

The final district that we discuss is Pudong New Area, a financial hub that has developed rapidly in the last few decades. This area, which is located across the Huangpu River from European concession territories and the old city of Shanghai, was initially developed only modestly before the late 20th century. In the 1990s, the Chinese government set up a Special Economic Zone in Pudong New Area \cite{sang1993}, and this district now has some of Shanghai's most famous skyscrapers. The PD of our street map of Pudong New Area [see Figure~\ref{fig:shanghai_pds}(c)] has several 1D features with death times that are larger than 20, indicating the existence of large blocks. We also observe several features with early and moderate death times; these correspond to a few small blocks on the map. For example, there appears to be a small traffic circle towards the western part of the street map [see Figure \ref{fig:streetmaps}(c)]. The large blocks are indicative of modern styles of urban planning, with large blocks laid out along grids. Although these blocks are much larger than those in the street maps of the other two regions, many of them are not rectangular, so we again observe several features with late birth times.

%%%%%%

\section{Case Study: Hotspots of COVID-19 Infections}
\label{sec4}

The spread of coronavirus disease 2019 (COVID-19), which is caused by severe acute respiratory syndrome coronavirus 2 (SARS-CoV-2), has resulted in a global pandemic \cite{WHO}. Modeling the spread of COVID-19 is an important and complicated task \cite{vesp2020}, in part because of the spatial heterogeneity in how it spreads. 

TDA can be useful for the analysis of spreading phenomena. For example, PH has been used previously in epidemiological applications to forecast the spread of Zika \cite{zika} and to analyze the Watts threshold model of a contagion on noisy geometric networks \cite{contagion}. PH provides a different perspective than the many spatiotemporal forecasting models that have been developed for COVID-19 without TDA (see, e.g., \cite{yuliagel, highresolution}). 

\begin{figure}
    \centering
    \subfloat[LA Neighborhoods]{\includegraphics[width = .45\textwidth]{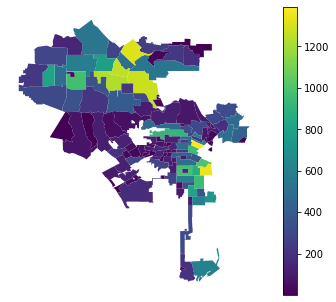}}
    \hspace{10mm}
    \subfloat[California Counties]{\includegraphics[width = .45\textwidth]{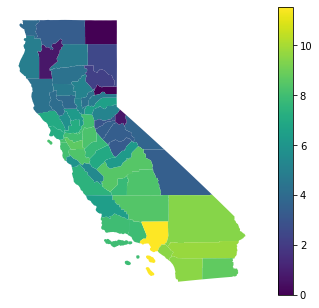}}
    \caption{Cumulative COVID-19 case counts on 30 June 2020 in (a) Los Angeles neighborhoods and (b) California counties. We plot the LA case counts on a linear scale and the California county case counts on a (natural) logarithmic scale.}
    \label{fig:datasets}
\end{figure}

In our case study, we use PH to analyze the spatial properties of the spread of COVID-19 in Los Angeles (LA) neighborhoods and California counties. In contrast to prior work, we use PH in a way that incorporates the underlying geographic structure and various spatial relationships. We consider two data sets. The first is a highly granular data set that consists of COVID-19 case counts in 136 LA neighborhoods on 30 June 2020. The second is a coarser data set that consists of case counts in the 58 counties of California on the same day \cite{CAdata}. For each data set, we also have geographic information in the form of a {\sc shapefile} \cite{CAshp, LAshp}. We visualize these data sets in Figure~\ref{fig:datasets}.

\begin{figure}
    \centering
    \subfloat[$M_0^{\text{LA}}$]{\includegraphics[width = .35\textwidth]{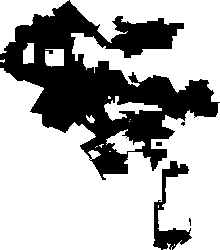}}
    \hspace{25mm}
    \subfloat[$M_0^{\text{CA}}$]{\includegraphics[width = .35\textwidth]{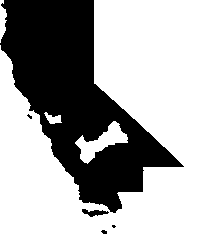}}
    \caption{Initial manifolds for the level-set filtrations that we construct from data of the spread of COVID-19. (a) The manifold $M_0^{\text{LA}}$ is an approximation of the manifold $M^{\text{LA}}$, which consists of the union of LA neighborhoods with fewer than $750$ cumulative cases on 30 June 2020. (b) The manifold $M_0^{\text{CA}}$ is an approximation of the surface $M^{\text{CA}}$, which consists of the union of California counties with fewer than $5,000$ cumulative cases on 30 June 2020.}
    \label{fig:M0}
\end{figure}

Let $M^{\text{LA}}$ denote the 2D manifold that consists of the union of LA neighborhoods with fewer than $750$ cumulative cases, and let $M^{CA}$ denote the union of California counties with fewer than $5,000$ cumulative cases. We approximate these manifolds by rasterizing the associated {\sc shapefiles} to obtain manifolds $M_0^{\text{LA}}$ and $M_0^{\text{CA}}$. We show $M_0^{\text{LA}}$ and $M_0^{\text{CA}}$ in Figure~\ref{fig:M0}. As we described in Section \ref{sec2}, we construct sequences of manifolds starting from $M_0^{\text{LA}}$ and $M_0^{\text{CA}}$ using level-set dynamics \eqref{level}. We then construct a level-set filtration for each of these sequences by imposing the manifolds in them on a triangulation of the plane.

\begin{figure}
    \centering
    \subfloat[LA neighborhoods]{\includegraphics[width = .47\textwidth]{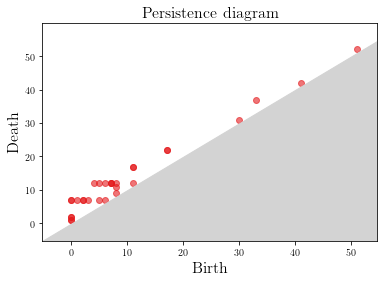}}
    \hspace{5mm}
    \subfloat[CA counties]{\includegraphics[width = .47\textwidth]{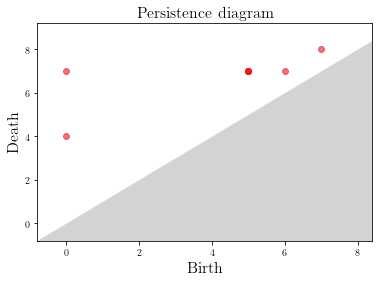}}
    \caption{The PDs of the level-set filtrations for COVID-19 cases in (a) LA and (b) California.}
    \label{fig:ph}
\end{figure}

In Figure~\ref{fig:ph}, we show the PDs that we compute for the 1D PH of our level-set complexes for the two data sets. These PDs can help us identify COVID-19 hotspots. We define a ``hotspot'' to be a collection of regions --- sets of neighborhoods in the LA data and sets of counties in the California data --- in which the case count is higher than in the surrounding area. This notion of a hotspot is analogous to the political ``islands'' that were studied using PH in \cite{feng2021}. Hotspots with a case count that is at least as large as the threshold (750 for LA neighborhoods and 5,000 for California counties) appear as holes in $M_0$, unless the hotspot is adjacent to the boundary of the map. The hotspots that are not adjacent to the boundary correspond to homology classes that are born at time $0$. 
Note that there is not a one-to-one correspondence between hotspots and homology classes that are born at time 0. Some of the homology classes that are born at $0$ are simply holes in the map (e.g., see Figure~\ref{fig:datasets}a), and hotspots that are adjacent to a boundary do not necessarily correspond to any homology class. Homology classes that are born after time $0$ usually reflect only the geography of the regions, although they sometimes correspond to hotspots on the boundary of the map (much like the homology classes that are created by city blocks with dead ends in the Shanghai street networks). The PDs reflect both the number of hotspots and the sizes of the hotspots.

%%%%%

\section{Conclusions}
\label{sec5}

In this chapter, we discussed the importance of incorporating spatial information into TDA when one uses it to study spatial systems. As case studies, we computed PH using a level-set construction of filtered simplicial complexes on two case studies: city street networks in Shanghai and hotspots in the spread of COVID-19 infections.

In our case study of street networks in Shanghai, we illustrated that PH can capture both topological and geometric properties of the organization of city streets. We also observed that the PHs of Shanghai's street networks reflect underlying differences in urban planning and organization. This suggests that topological tools can summarize information about how humans organize themselves in space, although further study is necessary to fully understand what types of spatial organization are amenable to TDA.

In our case study of the spread of COVID-19, we showed that one can use a level-set filtration to study the number and sizes of COVID-19 hotspots on both a granular level (by considering neighborhoods in Los Angeles) and a coarse level (by considering counties in California). We used only case counts in our computations, but one can also construct level-set filtrations for the death counts, hospitalization counts, or other quantities. The level-set filtration is flexible, but our approach has important limitations. For example, we only detected hotspots with a case count that is above some fixed threshold. This restricts us to measuring the severity of an outbreak based on its geographic area. One way to address this issue is by applying the level-set filtration after constructing a cartogram \cite{cartogram}, instead of directly from a {\sc shapefile}. Additionally, the level-set filtration is unable to detect hotspots that occur on the boundary of a map. Addressing these limitations is part of ongoing work.

Many spatial systems are also social in nature, and there are major challenges to overcome when studying such systems using TDA. In this chapter, we studied spatial systems, but it is important to point out that many spatial systems (including the examples in this chapter) reflect complicated social dynamics. For example, the Shanghai street networks have been shaped by social processes like colonial occupation and displacement of historical neighborhoods. Additionally, COVID-19 is known to disproportionately affect certain communities because of a confluence of social factors, including who is in prison \cite{prison2016}, where hospitals are located \cite{hospitals2019}, and so on. The interaction between social and spatial systems is complicated and inseparable, and intense work is necessary to connect approaches like TDA in spatial systems to the social factors at play.

%%%%%

%\begin{acknowledgement}

\section*{Acknowledgements}

We thank the Los Angeles County Department of Public Health for providing the LA County data on COVID-19, and we thank Federico Battiston and Giovanni Petri for the invitation to write this chapter. We acknowledge support from the National Science Foundation (grant number 1922952) through the Algorithms for Threat Detection (ATD) program. MAP also acknowledges support from the National Science Foundation (grant number DMS-2027438) through the RAPID program.

%\end{acknowledgement}

%%%%%

%\bibliographystyle{plain}
%\bibliography{PT2}

%%%%%%

%%%%%

\end{document}